\input harvmac
% +--------------------------------------------------------------------+
% |                                                                    |
% |                           TABLES.TEX                               |
% |                                                                    |
% |                     Ray F. Cowan  15-Feb-85                        |
% |                                                                    |
% |                       Princeton University                         |
% |                                                                    |
% |          Present Address:  Laboratory for Nuclear Science          |
% |                            M.I.T.                                  |
% |                            Cambridge, MA 02139                     |
% |                                                                    |
% |                   E-mail:  rfc@slacvm.slac.stanford.edu            |
% |                                                                    |
% |                                                                    |
% |                     Last Revision: 17-Apr-86                       |
% |                                                                    |
% |   Macros I find handy for making tables.  See TABLEDOC TEX for     |
% |   a longer description.  The token-counting macros are straight    |
% |   from the TeXbook's "Dirty Tricks" appendix.                      |
% |                                                                    |
% +--------------------------------------------------------------------+
%
\newbox\hdbox%
\newcount\hdrows%
\newcount\multispancount%
\newcount\ncase%
\newcount\ncols% This is the number of primary text columns in the table.
\newcount\nrows%
\newcount\nspan%
\newcount\ntemp%
\newdimen\hdsize%
\newdimen\newhdsize%
\newdimen\parasize%
\newdimen\spreadwidth%
\newdimen\thicksize%
\newdimen\thinsize%
\newdimen\tablewidth%
\newif\ifcentertables%
\newif\ifendsize%
\newif\iffirstrow%
\newif\iftableinfo%
\newtoks\dbt%
\newtoks\hdtks%
\newtoks\savetks%
\newtoks\tableLETtokens%
\newtoks\tabletokens%
\newtoks\widthspec%
%
%  Book-keeping stuff--see how often these macros are called.
%
%  MOD RFC 900221.
%  Removed usage logging:  it's too complicated under VM/XA.
%\immediate\write15{%
%CP SMSG GJMSINK TEXTABLE --> TABLE MACROS V. 851121 JOB = \jobname%
%}%
%
%  Turn on table diagnostics.
%
\tableinfotrue%
\catcode`\@=11%  Allows use of "@" in macro names, like PLAIN.TEX does.
%  Debugging aid.  Writes #1 on the
%                                    user's terminal and in the log file.
%
%  Define the \tstrut height, depth in terms of the x_height parameter.
%
\def\tstrut{\vrule height3.1ex depth1.2ex width0pt}%
\def\and{\char`\&}%  Allows us to get an `&' in the text.  This is the
%                    same as using the PLAIN TeX macro \&.
\def\tablerule{\noalign{\hrule height\thinsize depth0pt}}%
\thicksize=1.5pt%  Default thickness for fat rules.  The user should feel
%                  free to change this to his preference.
\thinsize=0.6pt%   Default thickness for thin rules.
\def\thickrule{\noalign{\hrule height\thicksize depth0pt}}%
\def\ctr#1{\hfil\ #1\hfil}%
%
%
%
%  Here are things for controlling the width of the finished table.
%
\tablewidth=-\maxdimen%
\spreadwidth=-\maxdimen%
\def\tabskipglue{0pt plus 1fil minus 1fil}%
%
%  Stuff for centering or not.
%
\centertablestrue%
%
%
%
%  \vctr vertically centers its argument in the row.
%
\parasize=4in%
\gdef\ARGS{########}%  Produces the correct number of #'s in the preamble
%                      by the time eveything is expanded and \halign sees
%                      it.
\gdef\headerARGS{####}%  Same as \ARGS, but used in \header macros.
\def\@mpersand{&}%  Allows us to get alignment tab characters later
%                   when we have made the character "&" an active macro.
{\catcode`\|=13%  Make |'s locally active.
\gdef\letbarzero{\let|0}%  Globally define a macro that allows us to
%                          keep active |'s from being expanded in edef's.
\gdef\letbartab{\def|{&&}}%
\gdef\letvbbar{\let\vb|}%
%  This \def will cause active |'s read by
%                            \ruledtable to be converted into double
%                            alignment tabs.
}%  End of locally active |'s.
{\catcode`\&=4%  Make these alignment tabs.
\def\ampskip{&\omit\hfil&}%  This local macro skips a vertical rule.
\catcode`\&=13%  Now make &'s into active macros.
\let&0%  This allows us to expand \ampskip in the next \xdef without
%        attempting to expand the & and getting an "undefined control
%        sequence" error.
\xdef\letampskip{\def&{\ampskip}}%
\gdef\letnovbamp{\let\novb&\let\tab&}
%  This will cause active &'s read by
%                                   \ruledtable to be converted into
%                                   double tabs and an \omit'ted \vrule.
}%  End of locally active &'s.
\def\begintable{%  Here we make |'s and &'s active characters so we can
%                  interpret them as macros.  Note that this action is
%                  true only until we encounter the matching \endgroup
%                  token later at the end of the \ruledtable macro.
   \begingroup%
   \catcode`\|=13\letbartab\letvbbar%
   \catcode`\&=13\letampskip\letnovbamp%
   \def\multispan##1{%  We must redefine \multispan to count the number
%                       of primary columns, not physical columns.
      \omit \mscount##1%
      \multiply\mscount\tw@\advance\mscount\m@ne%
      \loop\ifnum\mscount>\@ne \sp@n\repeat%
   }%  End of \multispan macro.
   \def\|{%
      &\omit\widevline&%
   }%
   \ruledtable%  Now we call \ruledtable to do the real work.
}%  End of \begintable macro.
\long\def\ruledtable#1\endtable{%
%
%  This macro reads in the user's data entries
%  and converts them into a ruled table.
%
%  Important note:  Many macros and parameters are re-defined here, and
%  these must be kept local to the table macros to avoid conflict with
%  their use outside of tables.  This is done by the \begingroup token
%  macro \begintable and the \endgroup token at the end of
%  this macro.
%
   \offinterlineskip%  Needed to make rules touch each other.
   \tabskip 0pt%  Needed for same reason as \offinterlineskip.
   \def\widevline{\vrule width\thicksize}%  Make outer \vrule's wider.
   \def\endrow{\@mpersand\omit\hfil\crnorm\@mpersand}%
   \def\crthick{\@mpersand\crnorm\thickrule\@mpersand}%
   \def\crthickneg##1{\@mpersand\crnorm\thickrule
          \noalign{{\skip0=##1\vskip-\skip0}}\@mpersand}%
   \def\crnorule{\@mpersand\crnorm\@mpersand}%
   \def\crnoruleneg##1{\@mpersand\crnorm
          \noalign{{\skip0=##1\vskip-\skip0}}\@mpersand}%
   \let\nr=\crnorule%  A shorter abbreviation.
   \def\endtable{\@mpersand\crnorm\thickrule}%
   \let\crnorm=\cr%  Allows us to use \cr for our own purposes.
%
%  Cause user-typed \cr's to follow a row with a \tablerule.
%
   \edef\cr{\@mpersand\crnorm\tablerule\@mpersand}%
   \def\crneg##1{\@mpersand\crnorm\tablerule
          \noalign{{\skip0=##1\vskip-\skip0}}\@mpersand}%
   \let\ctneg=\crthickneg
   \let\nrneg=\crnoruleneg
   \the\tableLETtokens%  Get the user's extra \let's, if any.
%
%  Put the data entries into a token register so we can scan through them
%  and see what the user is asking us to do.
%
   \tabletokens={&#1}%  We add an extra alignment tab to the beginning
%                       of the first row to allow for the first \vrule.
%
%  Now count how many rows are in the table and return the result in
%  count register \nrows; do the same for columns, and return that
%  in register \ncols.
%
   \countROWS\tabletokens\into\nrows%
   \countCOLS\tabletokens\into\ncols%
%
%  Now do a little arithmetic to convert the number of primary columns
%  into the number of physical columns that the alignment preamble must
%  prepare for;  similarly for rows.
%
   \advance\ncols by -1%
   \divide\ncols by 2%
   \advance\nrows by 1%
%
%  Tell the user how many rows and columns we found in his data, if he
%  wants to know.
%
   \iftableinfo %
      \immediate\write16{[Nrows=\the\nrows, Ncols=\the\ncols]}%
   \fi%
%
%  Now we actually go ahead and produce the table.
%
   \ifcentertables
      \ifhmode \par\fi%  Make sure we are in vertical mode.
      \line{%  The final table comes out as an \hbox of width the \hsize.
      \hss%  The final table will be centered left-to-right.
   \else %
      \hbox{%
   \fi
      \vbox{%
         \makePREAMBLE{\the\ncols}%  Generate the preamble.
         \edef\next{\preamble}%  This line and the next line force the
         \let\preamble=\next%    expansion of all \ARGS tokens into the
%                                appropriate number of #'s.
         \makeTABLE{\preamble}{\tabletokens}%  Go do the \halign here.
      }%  End of \vbox.
      \ifcentertables \hss}\else }\fi%  Finish the centering effect.
%                                       It is important that no spaces
%                                       follow the two `}' here.
%  }%  End of \line.
   \endgroup%  Return all local macros and parameters to their outside
%              values.
   \tablewidth=-\maxdimen%  Reset \tablewidth to normal.
   \spreadwidth=-\maxdimen% Same for \spreadwidth.
}%  End of macro \ruledtable.
\def\makeTABLE#1#2{%  Does an \halign for the \ruledtable macro.
   {%  Start of local parameter values.
   \let\ifmath0%     These macros would cause trouble if they were to be
   \let\header0%     expanded in the following \xdef; we \let them be
   \let\multispan0%  equal to a digit, because digits can't be expanded.
%
%  Set up the width specification here.
%
   \ncase=0%
   \ifdim\tablewidth>-\maxdimen \ncase=1\fi%
   \ifdim\spreadwidth>-\maxdimen \ncase=2\fi%
   \relax%  This \relax is absolutely necessary, without it the following
%           \ifcase will always take \ncase=0.
%
   \ifcase\ncase %
      \widthspec={}%
   \or %
      \widthspec=\expandafter{\expandafter t\expandafter o%
                 \the\tablewidth}%
   \else %
      \widthspec=\expandafter{\expandafter s\expandafter p\expandafter r%
                 \expandafter e\expandafter a\expandafter d%
                 \the\spreadwidth}%
   \fi %
%\out{Widthspec=[\the\widthspec]}%
%\out{Preamble=[\preamble]}%
   \xdef\next{%  We must force the preamble to be expanded BEFORE the
      \halign\the\widthspec{%
%        \halign is done;  this \edef\next{...}\next construction
%                does the trick.
      #1%  This is the preamble text.
      \noalign{\hrule height\thicksize depth0pt}%  Makes the top \hrule.
      \the#2\endtable%  This is the main body.
%
%     \noalign{\hrule height0.7pt depth0pt}%  Makes the last \hrule.
      }%  End of \halign.
   }%  End of \next.
   }%  End of local values.
   \next%  This \next must be outside of the local values, because now
%          we want those troublesome macros in the \let's above to have
%          their normal actions.
}%  End of macro \makeTABLE.
\def\makePREAMBLE#1{%  This macro generates the necessary preamble for a
%                      ruled table with #1 primary columns.
%                      (Primary columns means the number of columns NOT
%                       counting those used for vertical rules.)
   \ncols=#1%  Get the number of columns desired.
   \begingroup%  Start local parameter definitions.
   \let\ARGS=0%  This is the key to the whole thing; it prevents \ARGS
%                from being expanded in the following \edef's.
   \edef\xtp{\widevline\ARGS\tabskip\tabskipglue%
   &\ctr{\ARGS}\tstrut}%  A 1-column preamble.  Gets the sizing right.
   \advance\ncols by -1%  One column has been generated; decrement the
%                         counter.
   \loop%  Append as many further columns as needed to the preamble.
      \ifnum\ncols>0 %
      \advance\ncols by -1%
      \edef\xtp{\xtp&\vrule width\thinsize\ARGS&\ctr{\ARGS}}%
   \repeat
   \xdef\preamble{\xtp&\widevline\ARGS\tabskip0pt%
   \crnorm}%  Adds the last \vrule.
   \endgroup%  End of local parameters.
}%  End of macro \makePREAMBLE.
\def\countROWS#1\into#2{%  This counts the number of rows in #1 by
%                          looking for control sequences that end a row,
%                          e.g., \cr, \crthick, etc., and puts the result
%                          into count register #2.
   \let\countREGISTER=#2%
   \countREGISTER=0%
%  \out{In countROWS:  tokens are [\the#1]}%
   \expandafter\ROWcount\the#1\endcount%
}%
\def\ROWcount{%
   \afterassignment\subROWcount\let\next= %
}%
\def\subROWcount{%
%  \out{In subROWcount:  next is [\meaning\next]}%  Debugging aid.
   \ifx\next\endcount %
      \let\next=\relax%
   \else%
      \ncase=0%
      \ifx\next\cr %
         \global\advance\countREGISTER by 1%
         \ncase=0%
      \fi%
      \ifx\next\endrow %
         \global\advance\countREGISTER by 1%
         \ncase=0%
      \fi%
      \ifx\next\crthick %
         \global\advance\countREGISTER by 1%
         \ncase=0%
      \fi%
      \ifx\next\crnorule %
         \global\advance\countREGISTER by 1%
         \ncase=0%
      \fi%
      \ifx\next\crthickneg %
         \global\advance\countREGISTER by 1%
         \ncase=0%
      \fi%
      \ifx\next\crnoruleneg %
         \global\advance\countREGISTER by 1%
         \ncase=0%
      \fi%
      \ifx\next\crneg %
         \global\advance\countREGISTER by 1%
         \ncase=0%
      \fi%
      \ifx\next\header %
%     \out{In subROWcount:  next=header, ncase set=1}%
         \ncase=1%
      \fi%
%     \out{In subROWcount:  ncase is [\the\ncase]}%
      \relax%
      \ifcase\ncase %
         \let\next\ROWcount%
%        \out{subROWcount---> ncase=\the\ncase}%
      \or %
         \let\next\argROWskip%
%        \out{subROWcount---> ncase=\the\ncase}%
      \else %
      \fi%
   \fi%
%  \out{subROWcount---> NEXT=\meaning\next}%
   \next%
}%  End of macro \subROWcount.
\def\counthdROWS#1\into#2{%
\dvr{10}%
   \let\countREGISTER=#2%
   \countREGISTER=0%
\dvr{11}%
%  \out{In counthdROWS:  tokens are [\the#1]}%
\dvr{13}%
   \expandafter\hdROWcount\the#1\endcount%
\dvr{12}%
}%
\def\hdROWcount{%
   \afterassignment\subhdROWcount\let\next= %
}%
\def\subhdROWcount{%
%\out{In subhdROWcount:  next is [\meaning\next]}%
   \ifx\next\endcount %
      \let\next=\relax%
   \else%
      \ncase=0%
      \ifx\next\cr %
         \global\advance\countREGISTER by 1%
         \ncase=0%
      \fi%
      \ifx\next\endrow %
         \global\advance\countREGISTER by 1%
         \ncase=0%
      \fi%
      \ifx\next\crthick %
         \global\advance\countREGISTER by 1%
         \ncase=0%
      \fi%
      \ifx\next\crnorule %
         \global\advance\countREGISTER by 1%
         \ncase=0%
      \fi%
      \ifx\next\header %
%\out{In subhdROWcount:  next=header, ncase set=1}%
         \ncase=1%
      \fi%
%\out{In subhdROWcount:  ncase is [\the\ncase]}%
\relax%
      \ifcase\ncase %
         \let\next\hdROWcount%
%\out{subhdROWcount---> ncase=\the\ncase}%
      \or%
         \let\next\arghdROWskip%
%\out{subhdROWcount---> ncase=\the\ncase}%
      \else %
      \fi%
   \fi%
%\out{subhdROWcount---> NEXT=\meaning\next}%
   \next%
}%
{\catcode`\|=13\letbartab
\gdef\countCOLS#1\into#2{%
%  \out{In countCOLS:  tokens are [\the#1]}
   \let\countREGISTER=#2%
   \global\countREGISTER=0%
   \global\multispancount=0%
   \global\firstrowtrue
   \expandafter\COLcount\the#1\endcount%
   \global\advance\countREGISTER by 3%
   \global\advance\countREGISTER by -\multispancount
%  \out{countCOLS-->[\the\countREGISTER]}
}%
\gdef\COLcount{%
   \afterassignment\subCOLcount\let\next= %
}%
{\catcode`\&=13%
\gdef\subCOLcount{%
%\out{In subCOLcount: next is [\meaning\next]}
   \ifx\next\endcount %
      \let\next=\relax%
   \else%
      \ncase=0%
      \iffirstrow
         \ifx\next& %
            \global\advance\countREGISTER by 2%
            \ncase=0%
         \fi%
         \ifx\next\span %
            \global\advance\countREGISTER by 1%
            \ncase=0%
         \fi%
         \ifx\next| %
            \global\advance\countREGISTER by 2%
            \ncase=0%
         \fi
         \ifx\next\|
            \global\advance\countREGISTER by 2%
            \ncase=0%
         \fi
         \ifx\next\multispan
            \ncase=1%
            \global\advance\multispancount by 1%
         \fi
         \ifx\next\header
            \ncase=2%
         \fi
         \ifx\next\cr       \global\firstrowfalse \fi
         \ifx\next\endrow   \global\firstrowfalse \fi
         \ifx\next\crthick  \global\firstrowfalse \fi
         \ifx\next\crnorule \global\firstrowfalse \fi
         \ifx\next\crnoruleneg \global\firstrowfalse \fi
         \ifx\next\crthickneg  \global\firstrowfalse \fi
         \ifx\next\crneg       \global\firstrowfalse \fi
      \fi%  End of \iffirstrow.
\relax%\out{subCOL-->  ncase=[\the\ncase]}
% \out{subCOL-->  next=\meaning\next}
      \ifcase\ncase %
         \let\next\COLcount%
      \or %
         \let\next\spancount%
      \or %
         \let\next\argCOLskip%
      \else %
      \fi %
   \fi%
%  \out{subCOL-->  countREGISTER=[\the\countREGISTER]}
   \next%
}%
\gdef\argROWskip#1{%
%  Deletes the next balanced, undelimited argument from a
%                 token list.
% \out{---> Entering argROWskip <---}
% \out{In argROWskip:  deleted arg is [#1]}%
   \let\next\ROWcount \next%
}%  End of macro \argskip.
\gdef\arghdROWskip#1{%
%  Deletes the next balanced, undelimited argument from a
%                 token list.
% \out{---> Entering arghdROWskip <---}
% \out{In arghdROWskip:  deleted arg is [#1]}%
   \let\next\ROWcount \next%
}%  End of macro \arghdROWskip.
\gdef\argCOLskip#1{%
%  Deletes the next balanced, undelimited argument from a
%                 token list.
% \out{---> Entering argCOLskip <---}
% \out{In argCOLskip:  deleted arg is [#1]}%
   \let\next\COLcount \next%
}%  End of macro \argskip.
}%  End of active &'s.
}%  End of active |'s.
\def\spancount#1{%\out{spancount--->\meaning#1}
   \nspan=#1\multiply\nspan by 2\advance\nspan by -1%
   \global\advance \countREGISTER by \nspan
%  \out{number spancount--->\the\nspan; \the\countREGISTER}
   \let\next\COLcount \next}%
\def\dvr#1{\relax}%
% \omit\hfil%
% \parindent=0pt\hsize=1.1in\valign{%
% \vfil#\vfil&\vfil#\vfil\cr\hfil\hbox{\ Added to\ }\hfil&%
% \hfil\hbox{\ empty events\ }\hfil\cr}\hfil%
\def\header#1{%
\dvr{1}{\let\cr=\@mpersand%
\hdtks={#1}%
%\out{In header:  hdtks=[\the\hdtks]}%
\counthdROWS\hdtks\into\hdrows%
\advance\hdrows by 1%
\ifnum\hdrows=0 \hdrows=1 \fi%
%\out{In header:  Nhdrows=[\the\hdrows]}%
\dvr{5}\makehdPREAMBLE{\the\hdrows}%
%\out{In header:  headerpreamble=[\headerpreamble]}%
\dvr{6}\getHDdimen{#1}%
%\out{In header:  hdsize=[\the\hdsize]}%
%\striplastCR{#1}%
{\parindent=0pt\hsize=\hdsize{\let\ifmath0%
\xdef\next{\valign{\headerpreamble #1\crnorm}}}\dvr{7}\next\dvr{8}%
}%
}\dvr{2}}%  End of macro \header.
\def\makehdPREAMBLE#1{%This macro generates the necessary preamble for a
\dvr{3}%
%                      ruled table with \ncols primary columns.
%                      (Primary columns means the number of columns NOT
%                       counting those used for vertical rules.
\hdrows=#1%  Get the number of columns desired.
{%  Start local parameter definitions.
\let\headerARGS=0%
%  This is the key to the whole thing; it prevents \ARGS
\let\cr=\crnorm%
%                from being expanded in the followin \edef's.
\edef\xtp{\vfil\hfil\hbox{\headerARGS}\hfil\vfil}%
\advance\hdrows by -1%  One row has been generated; decrement the
%                         counter.
\loop%  Append as many further rows as needed to the preamble.
\ifnum\hdrows>0%
\advance\hdrows by -1%
\edef\xtp{\xtp&\vfil\hfil\hbox{\headerARGS}\hfil\vfil}%
\repeat%
\xdef\headerpreamble{\xtp\crcr}%
}%  End of local parameters.
\dvr{4}}%  End of \makehdPREAMBLE.
\def\getHDdimen#1{%
%\out{In getHDdimen:  Arg 1=[#1]}%
\hdsize=0pt%
\getsize#1\cr\end\cr%
}%  End of macro getHDdimen.
\def\getsize#1\cr{%
%\out{In getsize:  Arg 1=[#1]}%
%  Here we have to check arg#1 and see if the first token in #1 is an
%    \end; if so, we stop, else we check the width of arg#1.
%  We recall that each arg#1 will be terminated with a \cr token.
\endsizefalse\savetks={#1}%
%\out{In getsize:  the savetks = [\the\savetks]}%
\expandafter\lookend\the\savetks\cr%
%\out{In getsize:  ifendsize = [\meaning\ifendsize]}%
\relax \ifendsize \let\next\relax \else%
\setbox\hdbox=\hbox{#1}\newhdsize=1.0\wd\hdbox%
\ifdim\newhdsize>\hdsize \hdsize=\newhdsize \fi%
%\out{In getsize:  hdsize=[\the\hdsize]}%
%\out{In getsize:  newhdsize=[\the\newhdsize]}%
\let\next\getsize \fi%
\next%
}%
\def\lookend{\afterassignment\sublookend\let\looknext= }%
\def\sublookend{\relax%
%\out{In sublookend:  looknext = [\looknext]}%
\ifx\looknext\cr %
%\out{In sublooknext:  looknext=cr}%
\let\looknext\relax \else %
%\out{In sublooknext:  looknext/=cr}%
   \relax
   \ifx\looknext\end \global\endsizetrue \fi%
   \let\looknext=\lookend%
    \fi \looknext%
}%
%
%  Allow the user to make his own names for crthick, etc.
%
\def\tablelet#1{%
   \tableLETtokens=\expandafter{\the\tableLETtokens #1}%
}%
\catcode`\@=12%  Change @'s back to their normal category code.
%
%authordefs

\def \inparg{\leftskip = 40 pt\rightskip = 40pt}
\def \outparg{\leftskip = 0 pt\rightskip = 0pt}
\def\vev#1{\mathopen\langle #1\mathclose\rangle }
\thicksize=0.7pt
\thinsize=0.5pt
\def\ctr#1{\hfil $\,\,\,#1\,\,\,$ \hfil}  
\def\tstrut{\vrule height 2.7ex depth 1.0ex width 0pt}

\def\semi{;\hfil\break}

\def\frak#1#2{{\textstyle{{#1}\over{#2}}}}
\def\frakk#1#2{{{#1}\over{#2}}}

\def\hbar{{\overline h}{}}

\def\alphabar{\overline\alpha{}}
\def\betabar{\overline\beta{}}

\def\thetabar{{\overline \theta}}

\def\semi{;\hfil\break}
\def\npb{{Nucl.\ Phys.\ }{\bf B}}
\def\prd{{Phys.\ Rev.\ }{\bf D}}
\def\prl{Phys.\ Rev.\ Lett.\ }
\def\plb{{Phys.\ Lett.\ }{\bf B}}

\def\GeV{{\rm GeV}}

{\nopagenumbers
\line{\hfil LTH 592}
\line{\hfil hep-ph/0309165}
%\line{\hfil Revised Version}
\vskip .5in
\centerline{\titlefont Yukawa Textures and}
\centerline{\titlefont the mu-term}
\vskip 1in
\centerline{\bf\authorfont I.~Jack , D.R.T.~Jones\foot{address 
from Sept 1st 2003-31 Aug 2004: 
TH Division, CERN, 1211 Geneva 23, Switzerland} and R.~Wild}
\bigskip
\centerline{\it Dept. of Mathematical Sciences,
University of Liverpool, Liverpool L69 3BX, U.K.}
\vskip .3in

We show how with an  anomaly-free $U_1$, simple assumptions concerning 
the origin of Yukawa textures and the Higgs $\mu$-term 
lead to  the prediction of a new physics 
scale of $10^8\GeV$ and automatic conservation of baryon number. 

\Date{September 2003}}

\newsec{Introduction}

One of the most appealing approaches to the fermion mass 
hierarchy problem is provided by the Froggatt-Nielsen  (FN) mechanism
\ref\cdfr{C.D.~Froggatt and H.B.~Nielsen, \npb 147 (1979) 277
\semi 
M.~Leurer, Y.~Nir and N.~Seiberg,  \npb 398 (1993) 319
\semi 
P.~Ramond, R.G.~Roberts and  G.G.~Ross,
\npb 406 (1993) 19}.
According to FN, the hierarchy is produced from Yukawa textures 
produced by higher dimension terms involving MSSM singlet ``flavon'' 
fields $\theta$ via terms such as
$H_2 Q_i u^c_j (\frakk{\theta}{M_{\theta}})^{a_{ij}}$, 
where $M_{\theta}$ represents
the scale of new physics, and $a_{ij}= 0,1,2..$.
We consider here the case when the 
MSSM gauge group is extended by a single $U_1'$ 
group which is broken by $\vev \theta \neq 0$.
An exhaustive analysis of this general approach has been performed recently 
\ref\DreinerHW{
H.K.~Dreiner and M.~Thormeier,
hep-ph/0305270
%%CITATION = HEP-PH 0305270;%%
}\ 
by Dreiner and Thormeier (DT); this paper also contains a comprehensive 
list of references.
Our assumptions here differ from DT in two 
critical respects:

\item{*} We impose cancellation of all mixed $U'_1$ anomalies
\foot{For a recent account of how an anomaly free family-dependent  
$U'_1$ might be embedded in a replicated gauge group, see 
Ref.~\ref\ling{F.S.~Ling and P.~Ramond,
\prd 67 (2003) 115010}.} without 
invoking the Green-Schwarz 
mechanism \ref\grsch{M.~Green and J.~Schwarz, \plb 149 (1984) 117}. 

\item{*} 
We relax the assumption that there is only a single flavon field. 

{\noindent It might appear that our second assumption 
would rob us of most if not
all predictive power; we will show however, that there is a very  simple
naturalness criterion which results in a constrained  framework
resulting in definite predictions.  
This   arises as follows. 
Each Yukawa matrix $Y_{u,d,e}$ gains  its texture from a  {\it
particular\/} flavon,   $\theta_{u,d,e}$ with $U'_1$ charges $-Q_u$,
$-Q_d$,  and $-Q_e$, and we will choose $Q_u = 1$\foot{ 
We might want to assume that each flavon is 
accompanied by an oppositely charged $\thetabar$-partner; 
the simplest way to obtain a $U'_1$ $D$-flat direction, 
i.e. preventing the quadratic $D$-terms for the $U'_1$ 
from generating  
large masses for all the MSSM fields
\ref\ibross{L.E. Ib\'a\~nez and G.G. Ross, \plb 332 (1994) 100},
is to assume  the $\thetabar$s exist 
and have vevs approximately equal to the 
corresponding $\theta s$.
We will indicate  when this issue affects our discussion subsequently.}.
Our
naturalness criterion is simply that this state of affairs  arises by
virtue  of the charges of the fields, and is not imposed.   
Since we assume that the vevs  of the various  flavons are
approximately the same, with 
\eqn\lamdef{\vev{\theta_{u,d,e}}/M_{\theta} \approx \lambda \approx 0.22,}  
then if  we
want the $(11)$ entries of $Y_u$ and $Y_d$ to be  of order  $\lambda^8$
and $\lambda^4$ respectively, we could not have $Q_d = 2Q_u$ since
evidently were that so the $Y_u$ entry could be made  $O(\lambda^4)$ by
using $\theta_{d}$ instead of $\theta_{u}$. 
In imposing this criterion we will allow for possible flavon 
contributions to the Kahler potential.
The kinetic term for the quark doublets $Q$ will be, for example,  
$$
L = \Phi^*_i K_Q^{ij} \Phi_j 
$$
in superspace, where 
\eqn\Kdefn{
K_Q \sim \pmatrix{1 & \lambda^{k_1} & \lambda^{k_2}\cr 
\lambda^{k_1} & 1 & \lambda^{k_3}\cr \lambda^{k_2}& \lambda^{k_3}& 1\cr}
}
and $k_1 = k_2 + k_3$ (or a cyclic permutation).
Then we define $\Phi' = C_Q \Phi = D_QU_Q\Phi$, where $U_Q$ is the unitary 
matrix that diagonalises $K_Q$, so that $U_Q K_Q U_Q^{-1}= K_{\rm diag}$, and 
$D_Q$ is the  diagonal matrix whose 
entries are the square roots of the eigenvalues of $K_Q$.
Evidently 
\eqn\phidiag{
\Phi^*_i K_Q^{ij} \Phi_j = \Phi'^*_i  \Phi'_i,
}
and the Yukawa matrix $Y_u$, for example, will be replaced by 
$Y'_u = (C^{-1}_Q)^{T} Y_u C^{-1}_{u^c}$. 
It is important to realise that while the Yukawa terms are
holomorphic, so  that powers of $\theta^*_{u,d,e}$ cannot contribute to
them,  the Kahler terms are not. Note also, as remarked by DT, that 
the textures of $Y_u$ and $Y'_u$ may well differ, with, for example, 
texture zeroes being ``filled in". We, however, will restrict ourselves to 
cases when $Y_{u,d,e}$ already have our desired texture, and 
this texture is preserved by the canonicalisation.}

Thus far our analysis of the Kahler term 
mirrors that of DT. We differ from them 
in the following respect, however. We claim that quite generally the 
canonicalisation matrix $C$ can always be 
chosen (without fine-tuning) to  have the {\it same\/} texture as 
$K$. DT present an apparent counterexample, based on the matrix 
\eqn\DT{K = \pmatrix{1 & \lambda^{2} & \lambda^{4}\cr 
\lambda^{2} & 1 & \lambda^{2}\cr \lambda^{4}& \lambda^{2}& 1\cr}
}
but it is easy to construct a further unitary transformation that
reduces  their canonicalisation matrix to our claimed form; and a
unitary transformation obviously preserves the canonical kinetic form. 
Consider a simple $2\otimes 2$ example,
\eqn\exmp{K =\pmatrix{1&\lambda\cr \lambda & 1\cr}.}
This matrix is diagonalised by the transformation 
\eqn\Udef{
U = \frakk{1}{\sqrt{2}}\pmatrix{1&1\cr -1&1\cr},}
which is not close to the unit matrix; however the matrix
\eqn\CQdef{
C = U^{-1} D_Q U, \quad \hbox{where}\quad D_Q = \pmatrix{\sqrt{1+\lambda}&0\cr 
0&\sqrt{1-\lambda}\cr}} 
is a perfectly valid canonicalisation matrix, and takes the form
\eqn\Cseries{C = \pmatrix{1 + O(\lambda^2)&\frakk{\lambda}{2}+O(\lambda^3)\cr
\frakk{\lambda}{2}+O(\lambda^3)& 1 + O(\lambda^2)\cr},}
in accordance with our assertion. 

Let us turn now to a realistic example. 
Consider the ``Wolfenstein'' textures (see for example
Ref.~\ref\EIR{J.K.~Elwood, N.~Irges and P.~Ramond,
\prl 81 (1998) 5064}):
\eqn\textstan{
Y_u \sim \pmatrix{\lambda^{8}&\lambda^{5}&\lambda^{3}\cr  
\lambda^{7}&\lambda^{4}&\lambda^{2}\cr
\lambda^{5}&\lambda^{2}&1}, 
\quad
Y_{d} \sim \lambda^{\alpha_{d}}
\pmatrix{\lambda^{4}&\lambda^{3}&\lambda^{3}\cr  
\lambda^{3}&\lambda^{2}&\lambda^{2}\cr\lambda&1&1},
Y_{e} \sim \lambda^{\alpha_{e}}
\pmatrix{\lambda^{4}&\lambda^{3}&\lambda\cr  
\lambda^{3}&\lambda^{2}&1\cr\lambda^3&\lambda^{2}&1}
.}
The $Y_{u,d}$ textures lead to the Wolfenstein texture
for the CKM matrix, and appropriate hierarchies for the quark masses. 
There is considerable freedom in the choice of $Y_e$ 
texture; the above decision relates to the incorporation of neutrino masses, 
as will become clear anon. To avoid fine tuning of the 
leading order contributions  we would expect $\alpha_d \sim \alpha_e$
and $\tan\beta \sim \lambda^{\alpha_d -3}$; so we 
will restrict our attention to 
$ 3 \geq \alpha_{d,e} \geq 0$.  
Denoting the $U'_1$ charges of the multiplets 
$Q_i, L_i, u^c_i, d^c_i, e^c_i, 
H_1, H_2$ as $q_i, L_i, u_i, d_i, e_i, h_1, h_2$, it 
is easy to show that the  
mixed anomalies for $(SU_3)^2 U'_1$, $(SU_2)^2 U'_1$, 
$(U_1)^2 U'_1$ and   
$(U'_1)^2 U_1$ all cancel and the above textures are obtained 
if 
the following relations are satisfied:
\eqna\qecons$$\eqalignno{
Q_d &= 1 & \qecons a\cr
\Delta &= \alpha_d + 6   & \qecons b\cr 
Q_e &= 2 \alpha_d/(3 \alpha_e+6) & \qecons c\cr
u_1 &= -2\alpha_d/9 +16/3-2 h_2/3-e_1/3
+Q_e (10 + 3\alpha_e )/9
 & \qecons d\cr
e_1 &= - (116 - 12 Q_e \alpha_e + 32 \alpha_d - 24 h_2 - 40 Q_e + 
24 Q_e^2  + 20 Q_e^2  \alpha_e & \cr
& + 3 Q_e^2  \alpha_e^2  - 6 Q_e \alpha_e \alpha_d 
- 20 Q_e \alpha_d + 4 \alpha_d^2  - 4 \alpha_d h_2)/(2(\alpha_d+6)).
& \qecons e\cr}$$
Here $\Delta = h_1 + h_2$. We have not substituted 
for $Q_e$ in Eq.\qecons{e}\ and for $e_1$ and $Q_e$ in Eq.\qecons{d}\ 
because the resulting expressions are unwieldy. Note that $Q_d = Q_u$ 
so we only need two flavons at this stage. 
All the remaining charges are determined in terms of
$h_2$, $\alpha_e$ and $ \alpha_d$.

Let us now discuss the issue of {\it naturalness\/} we described above 
(ignoring at first the Kahler potential). 
Our system will be unnatural if there 
are solutions for $\alpha,\beta  \in \{0,1,2,3\cdots\}$ 
to any of the  following system of equations:
\eqna\natext$$\eqalignno{
\alpha  + \beta Q_e 
&= 8, \quad\quad\quad\quad\quad  \alpha + \beta \leq 7 & \natext a\cr
\alpha  + \beta Q_e 
&= (4+\alpha_d), \quad \quad \alpha + \beta \leq 3+\alpha_d   & \natext b\cr 
\alpha  + \beta Q_e 
&= (4+\alpha_e)Q_e,  \quad \alpha + \beta \leq 3+\alpha_e   & \natext c\cr
}$$
Note that for $\alpha_d \leq 4$ 
all the solutions to Eq.~\natext{b}\ are solutions 
to  Eq.~\natext{a}.
For any particular choice of $\alpha_{d,e}$ it is straightforward to 
classify the unnatural 
solutions for $Q_e$. Thus from Eq.~\natext{a}\ we obtain unnatural $Q_e$ values
\eqn\texts{
8,7,6,5,4,3,2,\frak{3}{2},\frak{5}{2},\frak{7}{2},\frak{4}{3},\frak{5}{3},
\frak{7}{3},\frak{8}{3},\frak{5}{4},\frak{7}{4},\frak{6}{5},\frak{7}{5},
\frak{8}{5},\frak{7}{6},\frak{8}{7},} 
while with, for example, $\alpha_e = 1$, $\alpha_d \leq 4$ we also have 
from Eq.~\natext{c} the additional unnatural values
\eqn\textb{
\frak{1}{2},\frak{1}{3},\frak{2}{3},\frak{1}{4},
\frak{3}{4},\frak{1}{5},\frak{2}{5},\frak{3}{5},\frak{4}{5}.}
If we were to assume the existence of $\thetabar$ flavon partners
(with similar vevs) then Eq.~\natext{a}, for example, would be replaced 
by 
\eqn\thbares{
\alpha -\alphabar + (\beta - \betabar)Q_e  = 8, \quad 
\alpha +\alphabar + \beta + \betabar \leq 7.} In that case an additional 
set of $Q_e$ values would be unnatural: Eq.~\texts\ would now also include 
the set\eqn\barset{
9,10,11,12,13,14,
\frak{9}{2},\frak{11}{2},\frak{13}{2},
\frak{10}{3},\frak{11}{3},\frak{9}{4},\frak{11}{4},\frak{9}{5}}
and Eq.~\textb\ the set \eqn\barsetb{
\frak{1}{6},\frak{1}{7},\frak{2}{7},\frak{1}{8},} and 
the corresponding negative charge would also be 
unnatural in every case in Eqs.~\texts, \textb, \barset, \barsetb.

Note that for $\alpha_d=\alpha_e=1$ we have from Eq.~\qecons{c}\ that 
$Q_e = 2/9$, which value appears in none of 
Eqs.~\texts, \textb, \barset, \barsetb. 
It is easy to establish that the possibilities 
$(\alpha_d,\alpha_e) = (1,0), (1,2), (2,0), (2,2),(3,1),(3,2),(3,3)$ 
are all unnatural, while  $(\alpha_d,\alpha_e) = (1,1), (1,3), 
(2,1),(2,3), (3,0)$ are natural. This conclusion continues to hold if 
we take into account the $\thetabar$-flavons. Note that $(3,0)$ gives 
$Q_e = 1$ so in this case we could have a single flavon; however 
(since $m_b > m_{\tau}$) this would manifestly require fine-tuning\DreinerHW. 
If we restrict to $\alpha_d \leq \alpha_e$, 
then we have three possible solutions.  
In what follows we will concentrate on $\alpha_d = 2$, $\alpha_e = 3$.

Turning to the Kahler terms, one sees easily that since $q_2 = q_1 -1, 
L_1 = L_2 + Q_e$ etc.,
we have 
\eqn\textsk{
K_{Q,e^c} \sim
\pmatrix{1&\lambda&\lambda^3\cr 
\lambda&1&\lambda^2\cr\lambda^3&\lambda^2&1},
K_{u^c} \sim \pmatrix{1&\lambda^{3}&\lambda^{5}\cr  
\lambda^{3}&1&\lambda^{2}\cr
\lambda^{5}&\lambda^{2}&1}, 
\quad
K_{d^c, L} \sim 
\pmatrix{1&\lambda&\lambda\cr  
\lambda&1&1\cr\lambda&1&1},}  providing the Kahler textures 
are generated by $\theta_u$ for $Q, u^c, d^c$ and by $\theta_e$ for 
$L, e^c$, and that in each case  only 
one flavon can contribute. For this to be natural we
must exclude solutions to (once again for 
$\alpha,\beta,\alphabar, \betabar  \in \{0,1,2,3\cdots\}$)  
\eqna\katext$$\eqalignno{
\alpha -\alphabar + (\beta - \betabar)Q_e  &= 5, \quad\quad 
\alpha +\alphabar + \beta + \betabar \leq 4 &\katext a\cr
\alpha -\alphabar + (\beta - \betabar)Q_e  &= 3Q_e, \quad 
\alpha +\alphabar + \beta + \betabar \leq 2 &\katext b\cr}$$

%
%\pm \alpha  \pm  \beta Q_e &= 5 \quad\quad  \alpha + \beta \leq 4 & 
%\katext a\cr
%\pm \alpha  \pm \beta Q_e &=  3 Q_e \quad \alpha + \beta \leq 2   
%& \katext b\cr 
%}$$
%
Thus for example $Q_e = -2$ is now also seen to be unnatural (this would,
of course be unnatural in any case if we were assuming the existence 
of $\thetabar$-partners). 
Note, however, that our anomaly cancellation conditions preclude $Q_e < 0$. 

It is easy to verify that, as we asserted earlier, the canonicalisation 
matrices corresponding to all the $K$-matrices in Eq.~\textsk\ have 
precisely the same texture as the corresponding $K$-matrix, and that 
the engendered transformations preserve the form of the textures 
in Eq.~\textstan.

We turn now to the Higgs $\mu$-term. If we suppose that it is 
generated in the same way as the Yukawa textures, that is via a term 
of the form $M_{\mu }H_1H_2 \lambda^{a_{\mu}},$ can we place any constraint on 
$a_{\mu}$?
Clearly we have $a_{\mu} = \alpha_{\mu} + \beta_{\mu}$ where 
\eqn\deldef{
\Delta = \alpha_{\mu}Q_u + \beta_{\mu}Q_e.
}
Now in our example, we see from Eq.~\qecons{c}\  that to obtain
$Q_e > 1$ we would need $\alpha_d > 3\alpha_e/2 +3$, which 
would again be difficult to reconcile with the fact that $m_b > m_{\tau}$. 
So we may assume $Q_e < 1$, and hence manifestly the smallest 
attainable value of $a_{\mu}$ is obtained for $\beta_{\mu} = 0$ and 
is $a_{\mu} = \Delta = \alpha_d + 6$. So in our favoured case 
$(\alpha_{d}, \alpha_e) = (2,3)$ we have $a_{\mu} = 8$ corresponding to 
$M_{\mu} \approx 10^8\GeV$ if we set $\mu = 500\GeV$. This conclusion 
is not altered if we assume the existence of $\thetabar$ flavon 
partners.

For $\alpha_{d,e} = (2,3)$ we list the various $U'_1$ charges in Table 1. 
An immediate consequence is that  the dimension 3,4 
$R$-parity violating operators of the form
$ LH_2, LLe^c, QLd^c, u^c d^c d^c, L^*H_1$ are all forbidden, 
not only in the sense that they are not  $U'_1$ invariant, but also 
in that they cannot be flavon generated: for example 
$L_1 + h_2 = 1223/300$, which manifestly can not be produced 
by a linear combination of $Q_u=1$ and $Q_e=4/15$.

\vskip3em
\vbox{

\begintable
Q_1      | Q_2           | Q_3     | L_1
        | L_{2,3}    | H_1   \cr
 \frakk{h_2}{3}    -\frakk{2309}{2700} |
\frakk{h_2}{3}-\frakk{5009}{2700}|\frakk{h_2}{3} -\frakk{10409}{2700}
| \frakk{1223}{300}- h_2| \frakk{381}{100}- h_2| 8 - h_2
\endtable
 
\vskip1em
%\vbox{
\begintable
 u_1      | u_2      | u_3  | d_1 | d_{2,3}\cr
\frak{23909}{2700}-\frak{4}{3}h_2|\frak{15809}{2700}-\frak{4}{3}h_2
| \frak{10409}{2700}-\frak{4}{3}h_2        | 
\frak{2}{3}h_2-\frak{3091}{2700}
  |\frak{2}{3}h_2-\frak{5791}{2700}
 \endtable
\vskip1em
%\vbox{
\begintable
 e_1     | e_{2} |e_{3}     \cr
  2h_2 - \frak{1021}{100}|  2h_2 - \frak{3143}{100}
|  2h_2 - \frak{1101}{100}
\endtable

\bigskip
\inparg  
{\centerline {\it Table~1:\/ The $U_1{}'$ hypercharges.}}
\bigskip 
\outparg}

Let us now explore the economical possibility that $M_{\mu} \sim M_{\theta}$. 
The objection to this is the possibility of flavon-generated 
baryon and lepton number violation: we would prefer not to 
impose these symmetries. We have already seen that 
dangerous dimension 3,4 operators are absent; but with $M_{\theta}$
so low we must obviously also consider higher dimension operators 
such as (here we list $B$-violating operators only)
\foot{For a listing of holomorphic higher dimension operators 
see Ref.~\ref\GherghettaDV{
T.~Gherghetta, C.F.~Kolda and S.P.~Martin,
\npb 468 (1996) 37}}
$$
\hbox{dimension 5}: QQQL, QQQH_1, u^c u^c d^c e^c, QQd^{c*} 
% u^c e^c  d^{c*}, 
$$
or 
$$
\hbox{dimension 6}:   QQQQu^c,  d^c d^c d^cLL,  d^c d^c d^c L H_1, 
 u^c u^c u^c e^c e^c,  u^c d^c d^c L H_2,  u^c d^c d^c H_1 H_2 ,
$$
$$
QQQH^*_2, QQu^{c*}e^{c*},   Q  u^{c*}  d^{c*} L,  Q  u^{c*}  d^{c*} H_1,
 Q  u^{c*}  d^{c*}  H_2^*, Q  d^{c*}  d^{c*}  H_2, 
$$
$$
Q  d^{c*}  d^{c*}  H_1^*,
Q  d^{c*}  d^{c*} L^*, 
d^c d^c d^c  e^{c*}.
$$ 
In all cases, given Table 1, these operators are not $U'_1$ invariant;
moreover (like the $R$-parity violating dimension 3,4 operators)  
they cannot cannot be flavon generated. 
In fact because our scale of new physics is so (comparatively) low, 
$B$-violating operators with dimension up to at least 8 are potentially 
dangerous. The number of such operators is large  so we do not list them.
We have, however, verified that there are no operators 
through dimension 9 violating {\it either\/} $B$ or $L$ that can 
be generated by any combination of our two flavon charges. Remarkably enough,
this conclusion was reached by simply examining all $B,L$ violating 
operators with $U_1$ hypercharge zero, without worrying whether they 
are $SU_3 \otimes SU_2$ invariant: this set manifestly contains the 
genuine $SU_3 \otimes SU_2 \otimes U_1$ operators. 
We conclude therefore that with $\theta$-charges $Q_{u,d,e} = 1,1,\frak{4}{15}$ 
and a physics scale $M_{\theta} \sim 10^8\GeV$, we can explain 
the matter mass hierarchy, the CKM matrix texture, and the magnitude 
of the Higgs $\mu$-term. 
So $B$ and $L$-violation associated 
with $M_{\theta}$ are highly suppressed; but since this includes the 
dimension 5 operator $\chi_{ij} = H_2L_iH_2L_j$ associated 
(generally via the see-saw mechanism) with 
neutrino masses, at this stage we have no explanation for the origin of 
the neutrino masses. However, 
the matrix of charges corresponding to 
$\chi_{ij}$ is easily constructed:
\eqn\chicharges{
Q_{\chi} = \pmatrix{\frak{1223}{150}&\frak{1183}{150}&\frak{1183}{150}\cr
& & \cr \frak{1183}{150}&\frak{381}{50}&\frak{381}{50}\cr
& & \cr \frak{1183}{150}
&\frak{381}{50}&\frak{381}{50}}.}
Then it is easy to see that if we introduce one more flavon with charge 
$-Q_{\nu}$ such that $Q_{\nu} = \frak{13}{150}$, 
we obtain a neutrino mass matrix with 
texture 
\eqn\neuttex{M_{\nu} \sim \frakk{v_2^2}{M_{\theta}}
\lambda^{10}\pmatrix{\lambda^2&\lambda&\lambda\cr
\lambda&1&1\cr \lambda&1&1},}
where the $\lambda^{10}$ arises because $7 + 2.(4/15) + 13/150 = 381/50$.   
This texture, as shown in 
Ref.~\ref\ramond{F.S.~Ling and P.~Ramond,
\plb 543 (2002)  29}, is compatible 
with current knowledge of the neutrino spectrum and mixing angles, 
without excessive fine-tuning (for a recent review of neutrino 
mass patterns see Ref.~\ref\KingIW{
S.F.~King, hep-ph/0306095
%%CITATION = HEP-PH 0306095;%%
}).

Moreover, even with the introduction of this new flavon, it remains the 
case that $B$-violation remains suppressed to at 
least the dimension 9 level. Because of the $\lambda^{10}$ factor 
in Eq.~\neuttex, we are thus able to generate  
neutrino masses with the same scale, $M_{\theta}$, as 
both the Yukawa couplings and the Higgs $\mu$-term. 
It is easy to check from Eq.~\neuttex\ that the largest 
neutrino mass is compatible with the ``normal hierarchy" neutrino 
spectrum.

We turn now to an alternative texture form which we previously employed 
in the context of Anomaly Mediation
\ref\jjw{
I.~Jack, D.R.T.~Jones and R.~Wild,
\plb 535 (2002) 193}\ref\jj{I.~Jack and D.R.T.~Jones,
\npb 662 (2003) 63}:
\eqn\textsdem{
Y_u \sim \pmatrix{\lambda^{8}&\lambda^{4}&1\cr
\lambda^{8}&\lambda^{4}&1\cr
\lambda^{8}&\lambda^{4}&1}, 
\quad
Y_d, Y_e \sim \lambda^{\alpha_{d,e}}
\pmatrix{\lambda^{4}&\lambda^{2}&1\cr
\lambda^{4}&\lambda^{2}&1\cr
\lambda^{4}&\lambda^{2}&1}.}

The Kahler textures are now given by:
\eqn\textsdd{
K_{Q,L} \sim
\pmatrix{1&1&1\cr 
1&1&1\cr1&1&1},
K_{u^c} \sim \pmatrix{1&\lambda^4&\lambda^8\cr  
\lambda^4&1&\lambda^{4}\cr
\lambda^{8}&\lambda^{4}&1}, 
\quad
K_{d^c, e^c} \sim 
\pmatrix{1&\lambda^2&\lambda^4\cr  
\lambda^2&1&\lambda^2\cr\lambda^4&\lambda^2&1},}  providing the Kahler textures 
are generated by $\theta_u$ for $u^c$, $\theta_d$ for $d^c$ and 
by $\theta_e$ for $e^c$. Once again we have that canonicalisation does not 
alter the form of the textures.
Cancellation of mixed anomalies leads to the following results:
\eqna\quafisol$$\eqalignno{
Q_d &= \frakk{\Delta-4}{\beta_d} & \quafisol a\cr
Q_e &= \frakk{2 (\Delta-6)}{3\beta_e} & \quafisol b\cr
u_1 &=  (60 \beta_e - 6 h_2 \beta_e  - 3 e_1 \beta_e + 4 \Delta - 24)
/(9\beta_e)& \quafisol c\cr
e_1 &=  2 [144(2\beta_e^2 - \beta_e^2  \beta_d^2 + 2\beta_d^2) 
+  \Delta^2(18\beta_e^2 + 8\beta_d^2
- 3   \beta_e^2  \beta_d^2 
+ 6   \beta_e \beta_d^2)\cr&
+\Delta ( 9 h_2 \beta_e^2  \beta_d^2  
    - 96 \beta_d^2  
 - 144\beta_e^2 - 36\beta_e \beta_d^2  
  - 18\beta_e^2  \beta_d^2) 
]  /  (9 \beta_e^2  \beta_d^2  \Delta)
& \quafisol d\cr
}$$
where $\beta_{d,e} = \alpha_{d,e} + 2$. Here we have assumed $\Delta\neq
0$; for  $\Delta = 0$ we require instead of Eq.~\quafisol{d}\ that
\eqn\delzero{2( \beta_e^2 + \beta_d^2) =  \beta_e^2 \beta_d^2.}
It is easy to show that (using the fact that 
$\alpha_{d,e} \in \{0,1,2,3\cdots\}$) $\alpha_d = \alpha_e = 0$ 
is the only possible solution to Eq.~\delzero. 
This case was analysed in Ref.~\jj\ in the AMSB context. 
Unfortunately, however, since for 
$\alpha_d = \alpha_e = \Delta =0$, we have that $Q_u = Q_d = 1, Q_e = -2$, 
it is easy to show that this 
case is clearly unnatural when we take
into  account the Kahler textures (or introduce $\thetabar$-flavons), 
so from the 
point of view of the present  paper is unsatisfactory. 
It is interesting, therefore, that
in the  AMSB  context we again find ourselves driven to $\Delta\neq 0$,
and hence a texture-generated $\mu$-term. 

Reverting to $\Delta\neq 0$, it is straightforward to enumerate 
the unnatural flavon charge assignments in the same way 
as we did for the Wolfenstein texture. Thus  
Yukawa unnaturalness will follow given a solution to any of:
\eqna\natextd$$\eqalignno{
\alpha  +  \beta Q_d  + \gamma Q_e 
&= 8, \quad\quad\quad\quad\quad\  \alpha + \beta + \gamma \leq 7 
& \natextd a\cr
\alpha  + \beta Q_d  + \gamma Q_e  
&= (4+\alpha_d)Q_d, \quad \alpha + \beta + \gamma \leq 3 + \alpha_d   
& \natextd b\cr
\alpha  + \beta Q_d  + \gamma Q_e   
&= (4+\alpha_e)Q_e,  \quad \alpha + \beta + \gamma \leq 3 +\alpha_e  
& \natextd c\cr
}$$
(with once again an obvious generalisation if we assume there are 
$\thetabar$ flavon partners).
It is easy to show that, for example, for $\alpha_d = \alpha_e = 0$, 
the following values of $\Delta$ are unnatural due to Eq.~\natextd{a-c}:
$$
30, 27, 24, 21, 20, 18,16, 15, 14,13, 12, 11, 10, 9, 8,7,6,5,4,3,2,-12,  
$$
$$
\frak{9}{2},\frak{11}{2},\frak{13}{2},\frak{15}{2},\frak{17}{2},\frak{19}{2},
\frak{21}{2},\frak{27}{2},\frak{33}{2},\frak{14}{3},
\frak{16}{3},\frak{19}{3},
\frak{20}{3},\frak{22}{3},\frak{26}{3}, \frak{28}{3}, \frak{32}{3}, 
\frak{15}{4},\frak{27}{4},\frak{33}{4},\frak{39}{4},\frak{45}{4},
$$
$$
\frak{12}{5},\frak{18}{5},\frak{21}{5},\frak{24}{5},\frak{32}{5},
\frak{33}{5},\frak{34}{5}, 
\frak{36}{5},\frak{38}{5},\frak{39}{5},
\frak{42}{5},\frak{44}{5},\frak{48}{5},\frak{51}{5},\frak{54}{5}, \frak{66}{5},
\frak{72}{5},
$$
$$
\frak{24}{7},\frak{30}{7},
\frak{44}{7},\frak{48}{7},\frak{51}{7},\frak{54}{7},
\frak{57}{7},\frak{60}{7}, \frak{66}{7},  \frak{72}{7},\frak{78}{7},
$$
$$
\frak{57}{8},\frak{78}{11},\frak{84}{11},
\frak{90}{11},\frak{96}{11},  
\frak{102}{11}, \frak{108}{11}, \frak{96}{13},\frak{102}{13},
\frak{108}{13}, 
\frak{114}{13}, \frak{120}{13}, \frak{114}{17}, \frak{120}{17},\frak{132}{17},  
\frak{132}{19}.
$$
It is remarkable that although (unlike in the Wolfenstein case) 
$\Delta$ is a free parameter, 
we can still limit the mass scale associated with the Higgs $\mu$-term. 
This time we have  $a_{\mu} = \alpha_{\mu} + \beta_{\mu} + \gamma_{\mu}$ 
with
\eqn\deldef{
\Delta = \alpha_{\mu}Q_u + \beta_{\mu}Q_d + \gamma_{\mu}Q_e.}
Substituting for $Q_{u,d,e}$ we obtain
\eqn\deltdd{
\Delta\left[3\beta_d\beta_e -3\beta_{\mu}\beta_e-2 \gamma_{\mu}\beta_d\right]
= 3\left[\beta_d\beta_e\alpha_{\mu}-4\beta_e\beta_{\mu}-4\gamma_{\mu}\beta_d
\right]}
Now manifestly if we choose $\beta_{\mu}, \gamma_{\mu}$ so that 
\eqn\consd{3\beta_d\beta_e -3\beta_{\mu}\beta_e-2 \gamma_{\mu}\beta_d = 0}
then we will obtain (independent of $\Delta$) the result
\eqn\amures{
a_{\mu} = \frakk{\beta_d+4}{\beta_d}\beta_{\mu} + 
\frakk{\beta_e+4}{\beta_e}\gamma_{\mu},}
or using Eq.~\consd{
\eqn\amufinal{
a_{\mu} = \beta_d + 4 +\frakk{3\beta_e -2\beta_d + 4}{3\beta_e}\gamma_{\mu}}
whence, if $3\alpha_e \geq 2\alpha_d - 6$ (which is true 
given our assumption $\alpha_e\geq\alpha_d$), the dominant 
contribution to the $\mu$-term is obtained by taking $\gamma_{\mu} = 0$.  
It then follows that, independent of the choice of $\Delta$ 
or the other unconstrained charge, the  $\mu$-term once again cannot 
be suppressed by a power greater than $\lambda^{\alpha_d + 6}$.

It would be logical now to reconsider the above discussion 
for the case when the $\thetabar$-flavons are present, but 
we will omit this because this texture scenario has a serious problem
as follows. Examining  $B, L$ violating operators, 
one easily finds, (for arbitrary $\Delta$ and $\alpha_{d,e}$)
that there are a number of dangerous dimension 5 operators: most 
catastrophically  $u^c_1u^c_3d^c_2e^c_2$ has $U'_1$ charge 
zero and is hence suppressed only by a single power of $M_{\theta}$.
This happens both for $\Delta \neq 0$ and 
$\Delta = 0$.  
Although in this framework  the right-handed flavour rotation is 
suppressed\jj\ it would require considerable fine-tuning to suppress 
it sufficiently to prevent an unacceptable proton decay rate from 
this operator.

In conclusion: the generalisation to several flavon fields relaxes 
some of the constraints on the anomaly free FN scenario, 
but it remains predictive 
if we assume a common mass scale origin for the Yukawa textures  
and the $\mu$-term. The solution we have described is based on 
Eq.~\textstan, and predicts that $M_{\theta} \sim 10^8\GeV$. Other 
lepton textures (for example $Y_e \sim Y_d$) are also possible; 
however the choice made in Eq.~\textstan\ enables us to also 
accommodate neutrino masses, albeit by means of a somewhat bizarre 
choice for the neutrino flavon charge. 
These textures are not satisfactory for the AMSB 
scenario described in Ref.~\jj\ because of FCNC effects associated 
with the Fayet-Iliopoulos $D$-terms; the alternative textures which 
avoid this problem (Eq.~\textsdem) turn out to be unsatisfactory 
from the point of view of naturalness that we have taken here. 

\bigskip\centerline{{\bf Acknowledgements}}

DRTJ was  supported by a PPARC Senior Fellowship, and was 
visiting the Aspen Center 
for Physics while part of this work was done. We thank 
Graham Ross and Marc Thormeier 
for conversations.

\listrefs

\end